\documentclass{aa}  

\usepackage{graphicx}
\usepackage{txfonts}
\title{Migration of Jupiter mass planets in discs with laminar accretion flows}
\author{ Lega, E. \inst{1}, Morbidelli, A.\inst{1}, Nelson, R.P.\inst{2}, Ramos, X.S. \inst{3}, Crida, A.
\inst{1} , B\'ethune, W. \inst{4} and Batygin, K. \inst{5}}
 \institute{Universit\'e C\^ote d'Azur, Observatoire de la Côte d'Azur, CNRS, Laboratoire Lagrange UMR7293,  Boulevard de l'Observatoire, 06304 Nice Cedex 4, France. \and Astronomy Unit, School of Physics and Astronomy, Queen Mary University of London, London E1 4NS, UK. \and Niels Bohr International Academy, Niels Bohr Institute, Blegdamsvej 17, DK-2100 Copenhagen Ø, Denmark.
 \and Institut für Astronomie und Astrophysik, Universität Tübingen, Auf der Morgenstelle 10, 72076, Tübingen, Germany \and Division of Geological and Planetary Sciences, California Institute of Technology, 1200 E. California Blvd., Pasadena, CA 91125, USA }

\begin{document}

% \abstract{}{}{}{}{} 
% 5 {} token are mandatory
 
  \abstract
% context heading (optional)
 {Migration of giant planets in discs 
 with low viscosity has been studied recently. Results have shown that the proportionality between migration speed and the disc's viscosity is broken by the presence of  vortices that appear at the edges of the planet-induced gap. Under some conditions, this 'vortex-driven' migration can be very slow and eventually stops.  However, this result has been obtained for discs whose radial mass transport is too low (due to the small viscosity) to be compatible with the mass accretion rates that are typically observed for young stars. 
}
% aims heading (mandatory)
{Our goal is to investigate vortex-driven  migration in low-viscosity discs in the presence of radial advection of gas, as expected from angular momentum removal by magnetised disc winds. }
% methods heading (mandatory)
{We performed  three dimensional simulations using the grid-based code FARGOCA.  We mimicked the effects of a disc wind by applying a synthetic torque on a surface layer of the disc characterised by a prescribed column density $\Sigma_A$ so that it results in a disc accretion rate of $\dot M_A = 10^{-8}$~M$_{\sun}$/y. We have considered values of $\Sigma_A$ typical of the penetration depths of different ionising processes. Discs with this structure are called 'layered' and the layer where the torque is applied is denoted as 'active'.  We also consider the case of accretion focussed near the disc midplane to mimic transport properties induced by a large Hall effect or by weak Ohmic diffusion.}
% results heading (mandatory)
{We observe two migration phases: in the first phase, which is exhibited by all simulations, the migration of the planet is driven by the vortex and is directed inwards.
This phase ends when the vortex disappears after having opened a secondary gap, as is typically observed in vortex-driven migration. Migration in the second phase depends on the ability of the torque from the planet to block the accretion flow. When the flow is fast and unimpeded, corresponding to small $\Sigma_A$, migration is very slow, similar to when there is no accreting layer in the disc. When the accretion flow is completely blocked, migration is faster (typically ${\dot r_{\rm p}} \sim 12$~AU/My at 5 au) and the speed is controlled by the rate at which the accretion flow refills the gap behind the migrating planet. The transition between the two regimes, occurs at $\Sigma_A \sim 0.2$~g/cm$^2$ and 0.65 g/cm$^2$ for Jupiter or Saturn mass planets at 5.2 au, respectively.} 
% conclusions heading (optional), leave it empty if necessary 
   {The migration speed of a giant planet in a layered protoplanetary disc depends on the thickness of the accreting layer. The lack of large-scale migration apparently experienced by the majority of giant exoplanets can be explained if the accreting layer is sufficiently thin to allow unimpeded accretion through the disc.}

   \titlerunning{Giant planet migration in discs with laminar accretion flows.}
  \authorrunning{E. Lega et al.}

\maketitle
\section{Introduction}
{ Raw observations of extra-solar planets show two distinct  populations of giant planets. The first is composed of objects orbiting very close to their host star (known as 'hot-Jupiters')  and the second is characterised by planets with an orbital period larger than  100 days, the so-called warm and cold Jupiters.

%However, when taking into account observational biases,} it appears that most giant planets (known as "cold-Jupiters") are located  between 1 and several au from their host star { \citep{Butler2006, Udry2007,Cumming2008,Howard2010Sci,Mayor2011}}, while there are hints that their number decreases again farther out \citep{Mayor2011,Fernandes2019}.  
  From a theoretical standpoint, the observations of giant planets at distances larger than 0.5 au  from their host stars (or periods larger than 100 days) are difficult to understand. Giant planets are expected to have started to form at the snowline, the place where the condensation of water creates an over-density of solid material and the rapid formation of the first planetesimals 
 \citep{IdaGuillot, SO, Draz17, Draz18}. The snowline in  young discs is expected to be at 5-7~au \citep{2021A&A...648A.101L,2020ApJ...901..166V},
   thus the current location of the cold Jupiters implies that they migrated just a few au in the protoplanetary disc, at odds with classical theoretical expectations.

   The migration of a planet changes regime as the planet grows in mass. When the planet has a mass smaller than 10-20 Earth masses, it migrates in the Type I regime. Then, as the planet grows more massive and starts opening a gap in the disc, the migration evolves towards the Type II regime, which is valid when a deep gap encapsulates the giant planet’s orbit.
   
   This paper is part of a project focussing on the Type II regime. Type II
migration is an important phase since Jupiter mass planets
are expected to form in young discs on timescales of about
one million years and this migration mode can therefore become
the dominant one when acting over expected disc lifetimes
of a few Myr  \citep{2019A&A...622A.202J}. For discs with traditional values
of the viscosity parameter $\alpha$ \footnote{that is with $\alpha$ in the interval $[10^{-4},10^{-3}]$} a giant
planet needs to start Type-II migration at at least  10-20 AU to remain
beyond ~1 AU from the parent star after a My of migration \citep{ColemanNelson2014,Bitschetal2015,Voelkeletal2020,Robertetal18}.
Our goal is to understand under which conditions Type-II
migration can become slow enough to allow  giant planets to start at
the snowline (5-7 au) without overshooting the warm-Jupiter region after
a $\sim $ My of migration.}

  % The aim of this paper is to focus on the Type II regime to understand if it is required that the cold Jupiters formed at distances of 10 - 20~au, or if they could have formed close to the snowline and may have simply migrated slowly.
  % Type II migration is an important phase since  Jupiter mass planets are expected to form in young discs on timescales of about one million years and this migration mode can therefore become the dominant one when acting over expected disc lifetimes of a few Myr \citep{2019A&A...622A.202J}.}
    \par
   The classic view of the Type II regime is that migration has a speed comparable to the viscous radial motion of the gas, because the planet and gap have to migrate in concert \citep{Ward1997}.  This process  has been revisited recently \citep{Durmann and Kley 2015,Robertetal18}. Results have confirmed that the migration rate is proportional to the disc's viscosity for values of the $\alpha$ viscosity parameter larger than $10^{-4}$, though it is not exactly equal to the unperturbed drift speed of the disc. For these values of $\alpha$, Type II migration should reduce by a substantial fraction the orbital radius of a planet within the disc's lifetime \citep{Legaetal21}. \par
   While it was originally thought that viscosity in discs might arise from turbulence generated by the magneto-rotational instability (MRI; \citet{Balbus1991}), leading to $\alpha$ values larger than $10^{-4}$, it was later realised that the ionisation of the gas near the midplane of the disc is too weak to sustain the MRI \citep{1996ApJ...457..355G,Stone1996}. Even more recently, the inclusion of non-ideal MHD effects, such as ambipolar diffusion, led to the conclusion that  the coupling between the magnetic field and the gas should not make the disc turbulent even at its surface (see \citet{Turner2014} for a review). Thus, discs are probably much less viscous than previously thought.  
 Among the alternative mechanisms for generating turbulence, in purely hydrodynamic discs, the most promising is the Vertical Shear Instability (VSI hereafter \citet{Nelson2013,2014A&A...572A..77S}). However, the VSI should not be active at the disc midplane within a few astronomical units from the star because the cooling rates are too slow; a recent study in discs with increasing physical realism (purely hydrodynamics) provides values of $\alpha$ with an upper limit of $10^{-4}$ within 5 au \citep{Ziamprasetal2021}. These results show the relevance of studying  giant planet migration in very low viscosity discs.\par
In a recent paper \citep{Legaetal21}, we have considered disc models with very low viscosities where $\alpha$ has values in the interval $[0:10^{-4}]$.
We have shown that for $\alpha \leq 10^{-5}$ migration differs from classical Type-II migration in the sense that it is not proportional to the disc viscosity.
In discs with a vanishingly small viscosity, gap formation by a giant planet leads to the formation of a vortex at the outer edge of the gap. The evolution of the system then depends on the importance of self-gravity in the disc \citep{ZhuBaruteau16}. By comparing simulations with and without self-gravity, \citet{Legaetal21} identified a region of parameter space, mainly characterised by disc mass and orbital distance, where self-gravity can be neglected. Here, the planet remains on an almost circular orbit and inward migration occurs only as long as the disc can refill the gap left behind by the migrating planet, either due to gas diffusion caused by the presence of the vortex or because of the inward migration of the vortex itself due to its interaction with the disc. We have called this type of migration "vortex-driven migration".  This migration is very slow  and, in addition,  cannot continue indefinitely because eventually the vortex dissolves. Thus, this result potentially provides a promising way to explain the limited migration that most giant planets seem to have experienced. \par
 The \citet{Legaetal21} study, as well as the present one, is limited to disc masses and orbital distances from the star for which self-gravity can be neglected. Further studies, including self-gravity, will be necessary to study the migration of planets forming farther out in the disc, at distances where self-gravity cannot be neglected.
\par
 However, the discs considered in \citet{Legaetal21}, due to their small viscosities, have a  negligible radial mass transport,  incompatible with the mass accretion rates that are typically observed for young stars  ($\sim 10^{-8}M_\odot$/y, with a large, order of magnitude scatter around this value \citep{1998ApJ...495..385H,Manara2016}.
The aim of the present paper is to test if a similar migration pattern holds in more realistic discs with radial transport of gas towards the star consistent with typical accretion rates of young stars.
 Magnetically driven disc winds have been proposed as a mechanism to remove angular momentum from thin ionised surface layers of low viscosity protoplanetary discs, \citep{Suzuki09,BaiStone2013,Turner2014,2015ApJ...801...84G,Bethuneetal17} promoting in these layers a fast radial transport of gas towards the central star.
 However, using three dimensional non-ideal magneto-hydrodynamical simulations for planet migration studies would be prohibitive from the point of view of computational time; therefore in the present paper we use  simple hydrodynamical simulations in which we mimic the effects of magnetically driven winds. We generate radial transport of gas by applying a synthetic torque in a layer of finite thickness located either at the disc's surface or close to its midplane. The thickness of the layer is characterised by its column density, which is denoted by $\Sigma_A$ (indicating this is the ``active layer").
 A similar model has been introduced by \citet{McNallyetal2020}, who showed that the wind-driven accretion flow does not modify the migration  in the specific case of low mass planets. 
 
 In this work we extend that study to giant planet migration and explore the dependence of the results on the parameter $\Sigma_A$, looking for conditions that allow us to recover the slow vortex-driven migration mode unveiled in \citet{Legaetal21}.
 Additionally, we also consider the case where the torque due to the magnetic field induces a radial flow near the midplane. This is expected in the region of a disc within 10 au from the central star if the Hall effect is large and the magnetic field is aligned with the disc rotation axis \citep{BaiStone2013,Lesur14,Bethuneetal17}, or beyond 10 au if Ohmic resistivity is weak \citep{Lesur2021}. The case of radial flow induced near the midplane has been studied in the case of small mass planets embedded in  two dimensional inviscid discs \citep{McNallyetal2017}. The authors have shown that the dynamical corotation torque can slow down planet migration or even reverse it such that the planet runs away outwards.   This case has also been recently studied with a two dimensional model for the case of Saturn mass planets \citep{2020A&A...633A...4K} showing that, in the case of very strong wind, the planet may undergo rapid type III outward migration.  Clearly it is important to extend these studies to giant planet migration in three dimensional discs.
 \par
 The paper is structured as follows: in Section 2 we describe our physical model; in Section 3 we explain the modifications made to the code to generate radial gas transport and we provide the setup of simulations in Section 4. The migration of Jupiter-mass planets is described in Section \ref{sec:migr}; the case of a migrating Saturn mass planet is considered in Section \ref{sec:Saturn}. Conclusions and discussion are provided in Section \ref{sec:Conclusion}.

\section{Physical model}
The protoplanetary disc is treated as a three dimensional non self-gravitating gas whose motion is described by the Navier-Stokes equations.
We use spherical coordinates $(r,\varphi,\theta)$ 
where $r$ is the radial distance from the star, that is from the origin, $\varphi$ is the azimuthal coordinate
starting from the $x$-axis and $\theta$ the 
polar angle measured from the $z$-axis (the colatitude).  The midplane of the disc is at the equator
$\theta = {\pi \over 2}$.
We work in a  coordinate system which rotates with angular velocity:
$$\Omega_p^{0} = \sqrt {G(M_{\star}+m_p) \over {r_p(0)}^3}  $$
where $M_{\star}$ is the mass of the central star, { $G$ the gravitational constant}
and $r_p(0)$ is the initial distance to the star of  a planet of mass $m_p$, assumed to be on a circular orbit.
%\footnote {during planet migration
%$\Omega_p$  remains constant at the value corresponding to the %initial orbital radius of the planet.}.
The gravitational influence of the planet on the disc
 is modelled as in \cite{KBK09}
using a cubic-potential of the form: 

\begin{equation}
\Phi _p = \left\lbrace \begin{array}{ll}
-{m_pG\over d} &  d > \epsilon \\
-{m_pG\over d}f\left({d\over \epsilon}\right) & d\leq \epsilon   
\end{array} \right.
\label{cubic}
\end{equation}
where  $d$ is the distance from the disc element to the planet and  $\epsilon$ is the softening length;  our nominal value is $\epsilon = 0.4 R_H$ where  $R_H$ is the Hill radius: $R_H=r_p(m_p/3M_{\star})^{1/3}$ . The function $f$ is given by: 
 $f(x) = x^4-2x^3+2x $. \par
To the usual Navier-Stokes equations (see for example \cite{Lega14})
 we add an evolution equation for the internal energy per unit volume $e=\rho c_v T$ (where $\rho$
 and $T$ are respectively the volume density and the temperature of the disc and
$c_v$ is the specific heat at constant volume):
\begin{equation}
\label{energyeq}
{\partial e \over \partial t} + \nabla \cdot (e\vec v)
 =  -p\nabla \cdot \vec v  - c_v\rho \frac{T-T_0}{\tau_c}.
\end{equation}
Here $\tau_c$ is the cooling time ($\tau_c=2\pi/\Omega$,
 and $\Omega$ is the local orbital frequency); $T_0$ is the initial
temperature, defined as $T_0(r)=GM_*\mu h_0^2/R_{gas}r$, with $h_0$ being the disc
aspect ratio and $\mu$ the mean molecular weight ($\mu=2.3 g/mol $ for a standard solar mixture).\par
 The system of equations is closed defining  the pressure as:
$p=(\gamma-1)e$ 
with $\gamma=1.4$ the adiabatic index.
\par In short, we use
an adiabatic EoS, with exponential damping  of the temperature
perturbations to the initial temperature profile. We use Eq.\ref{energyeq} instead of  the simpler locally isothermal EoS
or the pure adiabatic model in order to avoid disc instabilities
like the vertical shear instability \citep{Nelson2013}.

\section{Radial transport of gas}
\label{sec:radialgastransp}
\subsection{Generating an accretion flow with an imposed torque: layered discs}
\label{radvel}
Instead of performing MHD simulations, which are prohibitive from the point of view of computer time, we mimic the removal of angular momentum from the surface layers of the disc by a magnetised wind, or the torque due to a horizontal magnetic field acting near the midplane, using a prescribed torque.\par
 We remark that  actual MHD winds would transport angular momentum towards the surface of the disc and eventually eject the gas, whereas in { purely hydrodynamical models we can only } consider the lower portion
of the wind where the magnetic torque causes accretion.
Moreover, disc winds in MHD simulations have a radially varying 'accretion efficiency', so that the radial mass accretion rate can vary  in steady state because the vertical outflow rate would remove the excess mass. In the present model we do not  have outflows, and therefore in order to set up a steady accretion disc  we generate radial transport of gas that corresponds to a radially constant mass accretion rate, in a layer with vertically integrated density:
$$\Sigma_A = \int _{z_a}^{\infty}\rho(z)dz.$$ 
The underlying idea is that this layer has a high enough ionisation fraction to be strongly coupled to the magnetic field, such that angular momentum removal is efficient. We describe this as the ``active layer''.  
Estimates of the ionisation rate as a function of $\Sigma_A$ can be found in \cite{Armitage19} and references therein. We treat the column density of the active layer as a free parameter and, according to  \cite{Armitage19} (see Fig.8), we test  $\Sigma_A = 0.1$, 1 and $10 \, g/cm^2$ as representative values of the column density of active layers with different ionisation rates. The mass flux towards the star in the active layer is denoted by $\dot M_A$. The radial velocity of the gas depends on $\dot M_A$ through the relation:
\begin{equation}
\label{vradtheo}
v_r= \frac  {\dot M_A} {2\pi r\Sigma_A}.
\end{equation}
The accretion flow is obtained by applying a synthetic torque, $\gamma_A$, to the corresponding layers.
Considering that the specific angular momentum is $L= \sqrt {GM_*r}$ , the
specific torque $\gamma_A$ required to generate the radial velocity $v_r$
is:
\begin{equation}
\gamma_A=\frac {dL} {dt}= \frac {v_r} {2}\sqrt {GM_*\over r}= \frac {\dot M_A} {4\pi\Sigma_A}\sqrt {GM_*\over r^3}.
\label{eq:torquea}
\end{equation}
In order to generate this torque we impose an acceleration in the
azimuthal direction ($\varphi$)
\begin{equation}
\frac {dv_{\varphi}}{dt}= \frac {\gamma_A} {r}= \frac  {\dot M_A} {4\pi\Sigma_A}\sqrt {GM_*\over r^5}.
\end{equation}
{ The acceleration in the azimuthal direction generates a radial velocity in the active layer of magnitude given by Eq.\ref{vradtheo}. We do not modify the other hydrodynamical variables; a mass flow will appear in the active layer as a consequence of the radial velocity generated by applying the synthetic torque.}\par
In order to have a smooth  transition between the active layer and the rest of the disc  where no torque is
applied, we multiply  $\gamma_A$ by a function $\mathcal{F}(g(z))$ with $\mathcal{F}$  the Fermi function:
\begin{equation}
\label{filter}
\mathcal{F}(x) = \frac {1} {1+\exp(-b(1-x))}.
\end{equation}
 The function $g(z)$ is the ratio between the disc column density
  $\Sigma(z) = \int_{z}^{\infty}\rho(s)ds$
 and $\Sigma_A$:
$g(z)=\Sigma(z)/\Sigma_A$;  $b$  is a parameter ($b=5$ in our simulations) regulating the steepness of the transition of $\mathcal{F}$ from  1 to 0 .
We remark that the Fermi filter satisfies: $\int_{0}^{\infty}\gamma_A \mathcal{F}(g(z)) \rho(z) dz=\gamma_A \Sigma_A$.\par
{In order to simulate radial accretion in the vicinity of the midplane we apply a synthetic torque to the column of gas with  $z\le H=h_0r$, while no torque is applied above. The transition between the torqued and untorqued regions is again implemented via the Fermi function with $g(z)=z/H$.}
\begin{table}
      % Give a unique label
%\vbox to100mm{\vfil
% For LaTeX tables use
%\begin{center}
%\begin{minipage}{100mm}
\begin{tabular}{|llll|}
\hline\noalign{\smallskip}
Name & $\Sigma_A (g/cm^2)$ & ${\dot M_A (M_{\odot}/y)}$ & ${\dot M_{D} (M_{\odot}/y)}$ \\
\noalign{\smallskip}\hline\noalign{\smallskip}
\hline\noalign{\smallskip}
$NW$  & 0  &  $0$ & $-8\,10^{-11}$ \\
$S01$  & 0.1  &  $-10^{-8}$ & $-8\,10^{-11}$ \\
$S1$  & 1  &  $-10^{-8}$ & $-8\,10^{-11}$  \\
$S10$  & 10  &  $-10^{-8}$ & $-8\,10^{-11}$\\
$Hin$  & 110  &  $-10^{-8}$ & $-8\,10^{-11}$\\
$Hout$  & 110  &  $10^{-8}$ & $-8\,10^{-11}$\\
\noalign{\smallskip}\hline
\end{tabular}
\caption{Simulations names and main parameters. The accretion  flow transported by the disc's active layer is labelled $\dot M_A$,  while $\dot M_{D}$ is the stellar accretion rate due to viscosity.  The minus sign indicates inflow. Notice the in the control simulation $Hout$ we apply a radial outflow in the vicinity of the disc midplane.}
\label{table:tab1} 
\end{table}

\subsection{Radial transport in classical viscous discs}
We compare migration in a disc with an active layer to that in a classical viscous $\alpha$-disc, and here we briefly review radial transport in a disc with the $\alpha$ viscosity prescription \citep{ShakuraSun1973}.
The viscosity $\nu$ is given by:
$$\nu(r)=\alpha c_s(r) h_0r,$$ with $c_s(r)$ the sound speed   defined as $c_s(r)=\sqrt{\gamma p/\rho}$ and $\alpha$ is a constant.
Angular momentum conservation in a steady-state disc links the radial transport (in the following $\dot M_{D}$)
to the disc surface density ($\Sigma(r)$) and the viscosity through:
\begin{equation}
  \dot M_{D}(r) = 3\pi \nu(r)\Sigma(r).
  \label{mdotnu}
\end{equation}
We assume $\alpha=10^{-5}$, because this value gave rise to vortex-driven migration in the simulations of \citet{Legaetal21} while making the simulations less sensitive to numerical resolution than those with $\alpha=0$. We note that we also assume $\alpha=10^{-5}$ in the layered disc models. 

\section{Simulation setup}

\begin{figure}
\includegraphics[height=4truecm,width=8truecm]{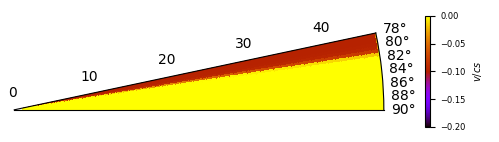}
\includegraphics[height=7truecm,width=8truecm]{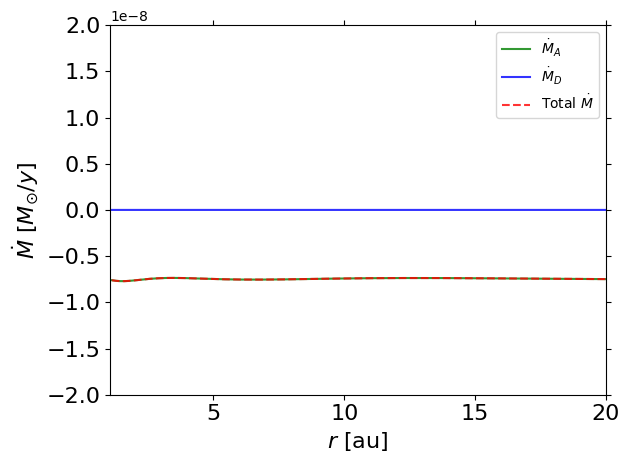}

\caption{Disc with radial 
  transport  $\dot M_{A}=10^{-8}M_{\odot}/y$  generated in the active layer with $\Sigma(z)<=0.1g/cm^2$. {\bf Top panel}: radial velocity normalised with respect to the sound speed; {\bf bottom panel}:
  total mass flow as a function of the distance from the star.
  The flow transported by the active layer is {$75\%$ of the nominal flow} $M_{A}=10^{-8}M_{\odot}/y$ (see text); the flow due to viscosity is clearly negligible with respect to  $\dot M_{A}$,  so that the total flow is transported by the active layer and the green and red curve are superposed
(the minus sign in the plot indicates inflow).}
\label{winddisc}
\end{figure}
Our three dimensional simulations are done with the code FARGOCA (FARGO with {\bf C}olatitude {\bf A}dded;
 \citet{Lega14})\footnote{The simulations presented in this paper have been obtained with a recently re-factorised version of the code that can be found at: https://gitlab.oca.eu/DISC/fargOCA}. The code is based on the  FARGO code \citep{Masset00} extended  to three dimensions. The fluid equations are solved using  a second order upwind scheme with a
time-explicit-implicit multi-step procedure.
The code is parallelised using a hybrid combination of MPI between the
nodes and OpenMP on shared memory multi-core processors. 
The code units are $G=M_*=1$, and the unit of distance $r_1=1$ is arbitrary when expressed in au.
The unit of time is therefore $r_1({\rm au})^{3/2}/(2\pi)\,{\rm yr}$. 
For the simulations in this paper we adopt the Sun-Jupiter distance as the unit of length in au: $r_1=5.2$. When presenting simulation results distances are expressed in au and time in years. The mass of the
planet is normalised with respect to the mass of the star and indicated by $q \equiv m_{p}/M_*$.\par
We consider discs of
initial aspect ratio $h_0= 0.05$ and radial domain extending from
$r_{min}\leq r \leq r_{max}$ with $r_{min}=1$ and $r_{max}=46$ (au) as in \citet{Legaetal21}.
In the vertical direction the discs have an opening angle of
$12^{\circ}$,  covering approximately 4.2 hydrostatic pressure scale heights from the midplane.  \par
We do not consider inclined planets and therefore, using the symmetry of the disc, we can simulate only the half-disc above the mid-plane (multiplying the resulting force by a factor of two to obtain the total force exerted on the planet). 
Mirror boundary conditions are applied at the midplane as in \citet{KBK09} and reflecting boundaries are applied at the disc surface.
In the radial direction we use the classical prescription \citep{ValBorro2006} of evanescent boundary conditions. \par

\begin{figure}
     \centering
     \includegraphics[height=6truecm,width=8truecm]{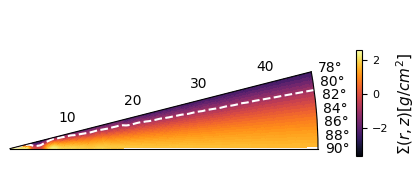}
     \caption{Vertically integrated surface density, $\Sigma(r,z) = \int ^{\infty}_z \rho(s)ds $, in simulation set S01 at time t=9481 y before the planet is released in the migration phase. The white dashed line corresponds to the $\Sigma_A = 0.1 g/cm^2$. The synthetic torque is applied above this line up to the disc surface. We remark that the formation of the gap implies that the synthetic torque is applied deep down in the disc.}
     \label{fig:integrated_density}
 \end{figure}
 
Discs have a  viscosity $\alpha=10^{-5}$ and a surface density
$\Sigma=\Sigma_0 (r/r_1)^{-1/2}$ with $\Sigma_0= 6.76\times10^{-4}$ in code units ($222\, \mbox{g\, cm}^{-2}$ at 5.2 au).\par
The exponent $-1/2$ in the surface-density radial profile ensures that the viscosity-driven mass flow $\dot M_{D}$ is independent of $r$, given that the disc is not flared.  The above values of viscosity and disc surface density  provide a mass transport: $\dot M_{D} =8\,10^{-11}M_{\odot} / y$ (from Eq.\ref{mdotnu}).\par
The subscript $D$ indicates that our discs are "dead" in the sense that a negligible amount of gas is transported due to  viscosity.
\par
The temperature profile is the same for all the discs: isothermal in the vertical direction with the midplane temperature scaling as $1/r$. The temperature is damped towards $T_0(r)\propto 1/r$  according to Eq. \ref{energyeq}.\par
Long run simulations in three dimensions are challenging in term of computational time,
therefore we have used a moderate resolution of $(N_r,N_{\theta},N_{\varphi})=(576,24,360)$  corresponding to about six gridcells per scale height and four gridcells per planet's Hill radius in the radial and azimuthal directions and height gridcells per planet's Hill radius in the vertical direction.
 Convergence of results with respect to resolution has been shown for the same set of parameters (in classical viscous discs) in \cite{Legaetal21} .\par
 In table \ref{table:tab1} we report  the main simulations parameters. { In the case of imposed radial flow in the vicinity of the midplane we consider also a control simulation with radial flow directed outwards: simulation set $Hout$. Although this simulation was mainly run to check the symmetry with respect to the $Hin$ case, we remark that in some cases magnetised discs have both accreting and decreting regions (\cite{Bethuneetal17}, Fig.22); thus the planet can indeed experience a local radial outflow of the disc.}
\begin{figure*}
   \includegraphics[width=16truecm,height=4truecm]{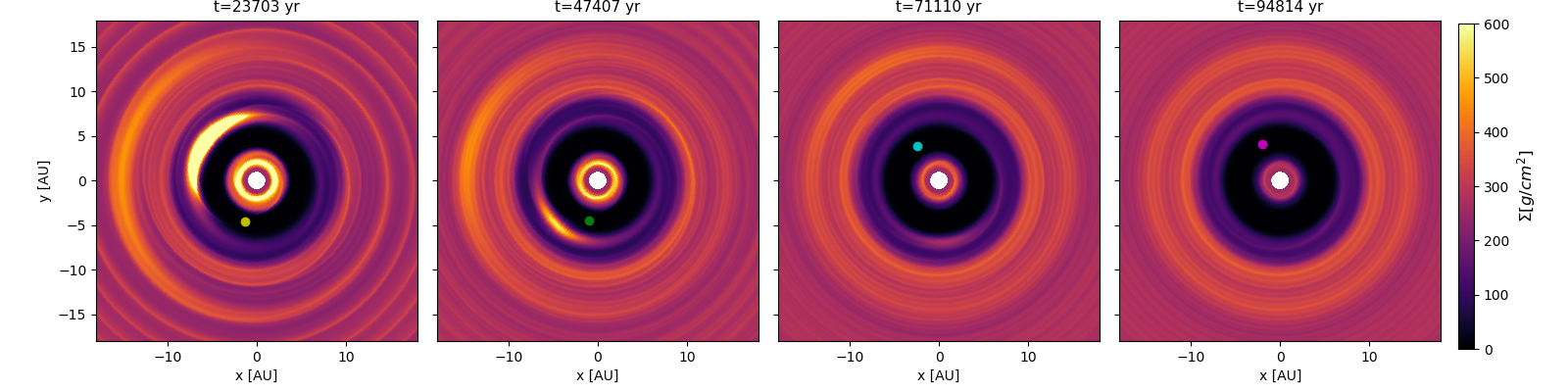}
   \includegraphics[width=16truecm,height=4truecm]{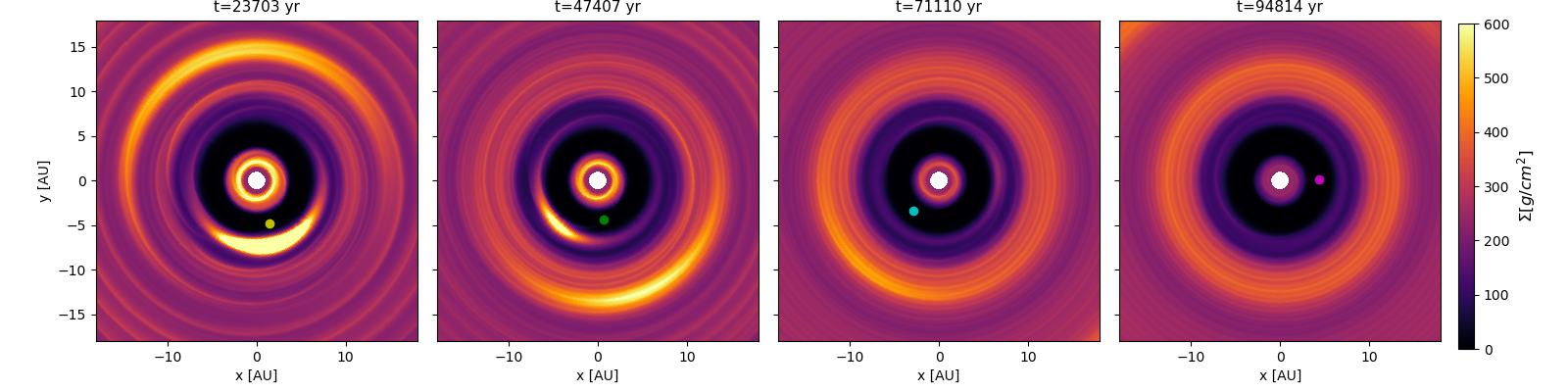}
   \includegraphics[width=16truecm,height=4truecm]{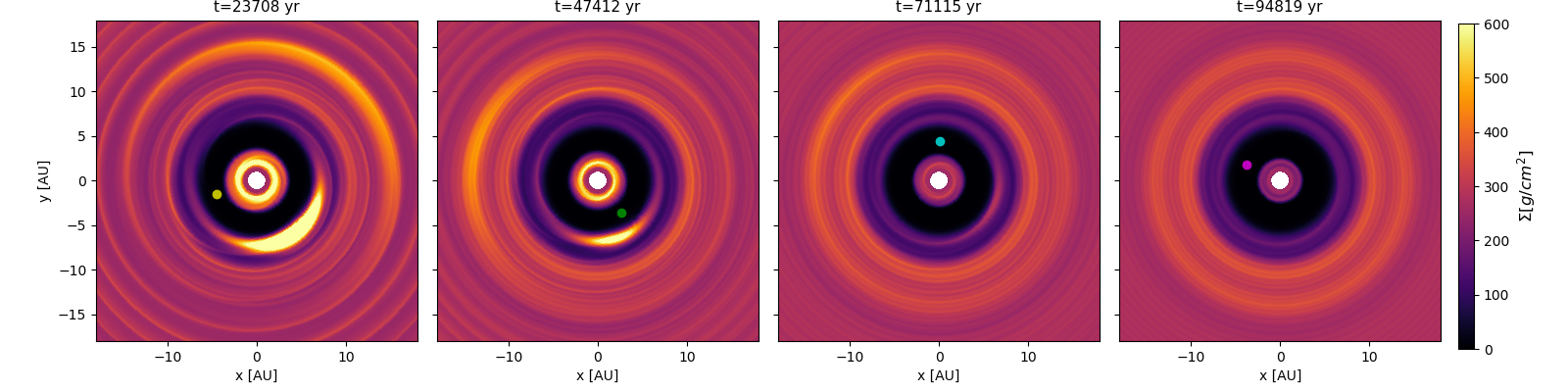}
   \includegraphics[width=16truecm,height=4truecm]{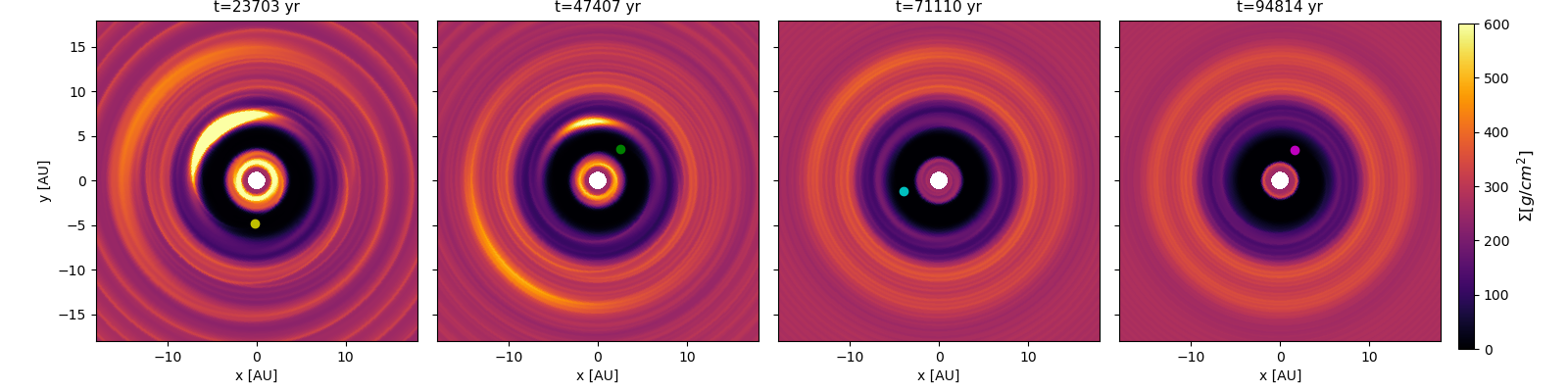}
   \includegraphics[width=16truecm,height=4truecm]{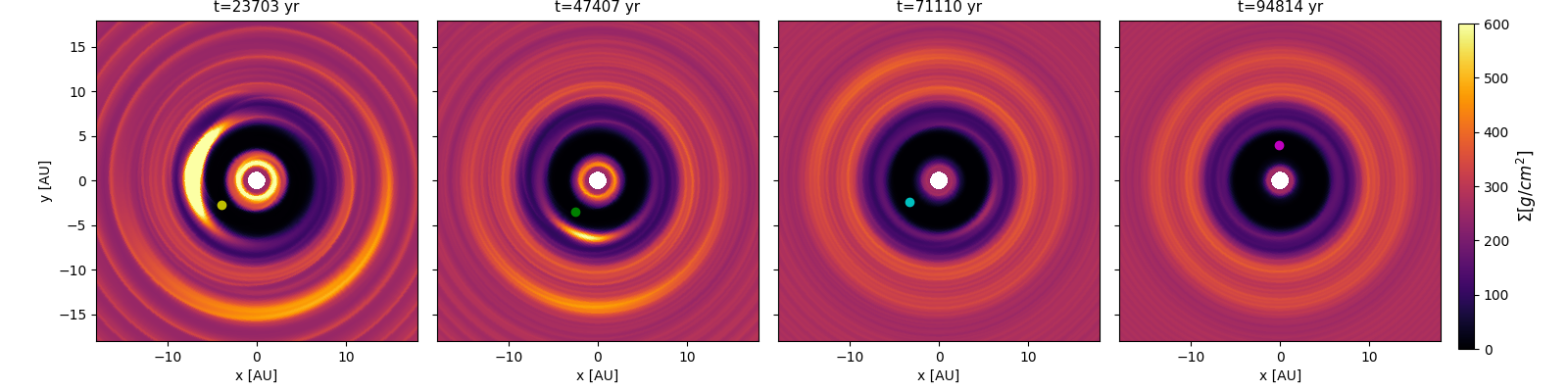}
   \includegraphics[width=16truecm,height=4truecm]{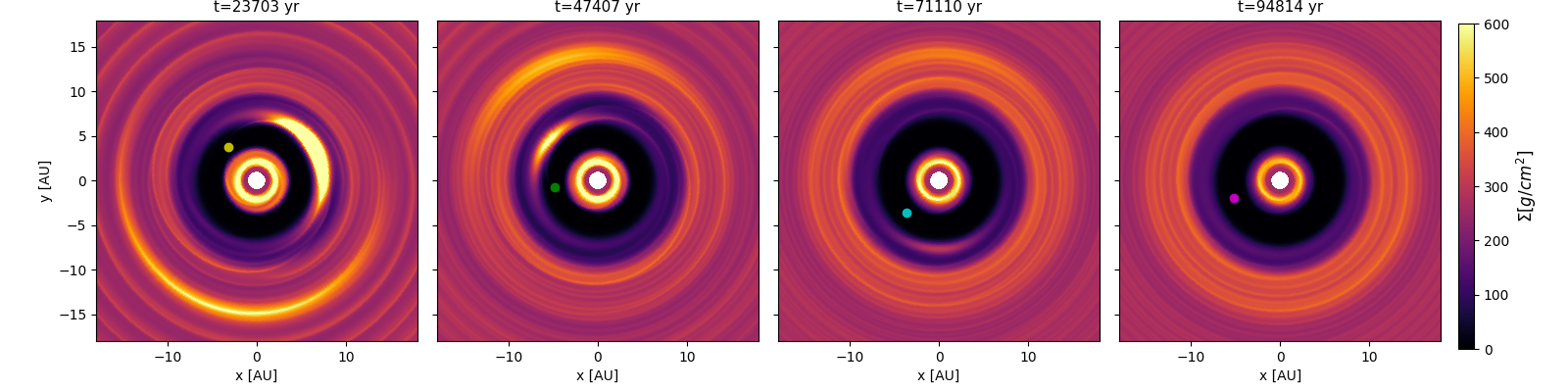}
      \caption{Contours of surface density (multiplied by $r^{1/2}$ to flatten the radial profile) for simulation sets $NW$, $S01$, $S1$, $S10$, $Hin$, $Hout$ (from top to bottom) with a migrating Jupiter-mass planet. The panels correspond to different times, reported on top of each panel. }
         \label{mig:surface} 
\end{figure*}

\subsection{Setup of a disc with radial transport in thin surface layers}
We show in this section the new disc setup corresponding to simulation set $S01$.
We assume  that the mass flow carried in the active layer is $\dot M_{A}= 10^{-8}M_{\odot} / y$ (each disc hemisphere carries $\dot M_{A}= 5\, 10^{-9}M_{\odot}/y$), as this is the typical accretion rate  observed in young stars \citep{1998ApJ...495..385H,Manara2016}.

Fig.\ref{winddisc}, top panel, shows in $(r,\theta)$ coordinates
 the radial velocity profile of the initial disc with no embedded planet.
 We clearly see that gas is transported through a thin surface layer.
In Fig.\ref{winddisc}, bottom panel, we show  the total mass flow in
$M_{\odot}/y$ transported across the disc.  We notice that the transport in the active layers  corresponds to  $75\%$ of the imposed flow  and that the remaining part of the disc doesn't provide any radial transport (or negligible with respect to $\dot M_{A}$ as expected). 
We notice that most of the gas radial flow comes from a very thin layer covered by few grid-cells near the transition of $\mathcal{F}$ from $1$ to $0$, { where the product $\rho(z) v_r$ has a sharp maximum because of the Gaussian shape of $\rho(z)$.  Hence the integrated mass flux is underestimated unless the transition is finely resolved.  
Precisely,} to reach  $95\%$ of the imposed flow in the active layers, a vertical resolution of $96$ grid cells is required.
 We stress that, although the flow is underestimated, this has no impact on the radial velocity which is the key parameter in our study.\par
In order to run simulations over $10^{5}$ years we keep our modest vertical resolution.
The discretisation effect on the flow transported to the star is the same for the 3 simulations sets with active layers (we have $80\%$ of the nominal flow for $S10$) and this makes the three cases comparable.
For the  simulations with transport in the vicinity of the midplane 
the gas carried by the active layers corresponds to the expected value with no dependence on resolution.
\par
  We remark that, although we are considering a purely hydrodynamic model, we capture the essential result of a disc with basically no transport near the midplane and fast accretion at its surface.  We notice for example  that the accretion layer in \cite{Bethuneetal17} is radially localised while we impose here accretion through the whole disc and that the fastest inwards radial velocity in pure MHD simulations is close to $c_s$ while it is about  $10\%$ of $c_s$ in our model S01 and are respectively $1\%$ and $0.1\%$ of $c_s$ in  models S1 and S10.

 \subsection{Introduction of the planet in the simulation}
 We introduce the planet on a fixed orbit and grow its mass up to its final Jupiter-mass value $q_J$  (or Saturn mass $q_S$, see Section~\ref{sec:Saturn}) in $T_{growth}=2\pi \times 600 $ according to:
 \begin{equation}
     q(t)= q_J\sin^2\left({{\pi }\over {2}} {{t} \over T_{growth}}\right).
     \label{eq:massgrowth}
 \end{equation}
 The simulation is then continued for an additional 200 orbits to stabilise the disc, before the planet is released in the migration runs { at $t=9481 y$}. In this way we avoid the excitation of instabilities that would arise if the planet were initialised with its final mass (see also \cite{2017MNRAS.466.3533H,2020MNRAS.491.5759H}). 
 
{  We have reported in Fig.\ref{fig:integrated_density} the vertically integrated surface density at {$t=9481 y$} for simulation $S01$. 
The white line delimits the region above which the synthetic torque is applied. We remark that, as described in Section \ref{sec:radialgastransp}, the synthetic torque is applied everywhere in the disc on a column of density $\Sigma_A$,  which means that in the gap region (in the vicinity of 5.2 au) 
the torque can be applied down to the midplane if the gas density in the gap is smaller than $\Sigma_A$
(see Section 5.1).  \par
 One may question whether it is realistic that the gas is ionised at all scale heights in the gap, which is the condition to be active. The answer depends on the incidence of the ionising radiation: whereas irradiation from the central star is almost tangent to the disc and should not penetrate directly into the gap down to the midplane, scattered x-rays and cosmic rays can reach the midplane if the column density is thin enough \citep{KimTurner2020}. Our model is certainly simplistic in this respect, given that we impose a thickness of the active layer without modelling self-consistently the value of this thickness, but it is worth noting that \citet{KimTurner2020} report that gas in the gap for a Jupiter or Saturn mass planet is expected to be sufficiently ionised to couple strongly to magnetic fields. \par In the present work we also consider that gap formation does not affect the magnetic flux distribution. Few papers have investigated magnetic fields inside gaps  \citep{CarballidoHyde2017,Zhuetal2013} and this challenging and important issue certainly deserves future dedicated studies. }

\section{Migration of a Jupiter mass planet}
\label{sec:migr}

\begin{figure}
\includegraphics[width=7.5truecm,height=7truecm]{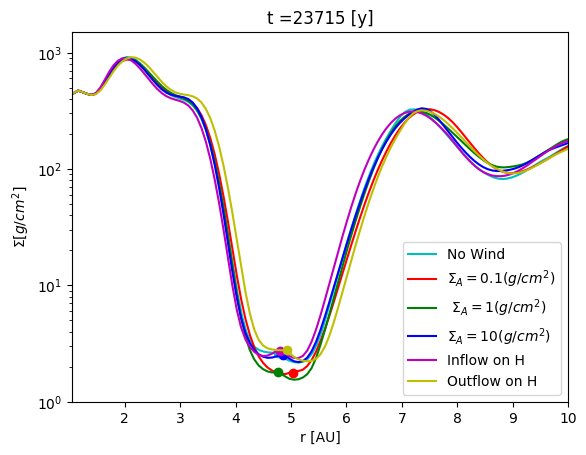}
   \includegraphics[width=7.5truecm,height=7truecm]{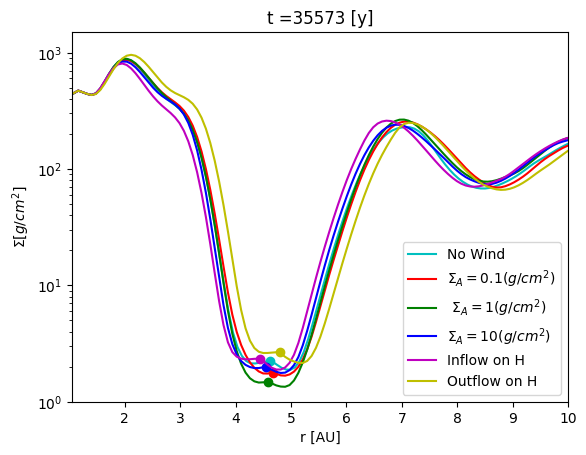}
   \caption{Azimuthally averaged radial surface density profiles  at two specific times (reported on top of the panels) within the rapid inward vortex-driven migration phase. No clear differences appear among the six simulations sets. }
\label{mig:sigma}
\end{figure}

We investigate the migration of the planet located initially at $r=5.2$~au for the six simulations  listed in Table\ref{table:tab1}.\par
The value of the Toomre $Q$ parameter in a disc with initial density $\Sigma_0=6.76 \times 10^{-4}$ at distance $r=5.2$~au from the star is much larger than 1 so that the effects of self-gravity on global disc evolution can be neglected \citep{Legaetal21}. \par
We recall that a giant planet opening a gap in the disc can migrate only at the speed at which the disc can readjust itself in order for planet and gap to migrate together \citep{Ward1997}. In viscous discs, the gas can displace radially on the viscous timescale, which sets the planet migration speed to be proportional to $\nu/r$.
In low viscosity discs the growth of a giant planet leads to the formation of a vortex  at the  pressure bump located at the outer edge of the gap. 
The vortex clearly appears in the leftmost panels of
Fig.\ref{mig:surface} with the same qualitative characteristics for the six simulations sets.
 The vortex migrates inwards and  spreads radially  until it  disappears (rightmost panels of Fig.\ref{mig:surface}). Vortex migration is due to its ability to exchange angular momentum with the disc, via the spiral density waves that it generates \citep{Paardekooperetal10}. Vortex migration is inwards because the vortex is located at the outer edge of the gap, so that the disc inside the vortex's orbit is strongly depleted and the outer disc dominates the interaction with the vortex.
 We have shown in \cite{Legaetal21} that the vortex transfers angular momentum to the disc beyond its orbit, carving a
secondary gap (second and third columns of Fig.\ref{mig:surface}). The secondary gap is also
clearly shown in Fig.\ref{mig:sigma}. Once this gap is formed, the vortex slows down, spreads radially and eventually dissipates completely, leaving the outer disc partially depleted near the planet's orbit. 
It is also worth noticing in Fig.\ref{mig:sigma}
that the gap depth and shape, in this first migration phase, is the same for all
the simulation sets, an indication that the vortex dominates the dynamics despite the existence of an imposed advection.

\begin{figure}
   \includegraphics[width=9truecm,height=6truecm]{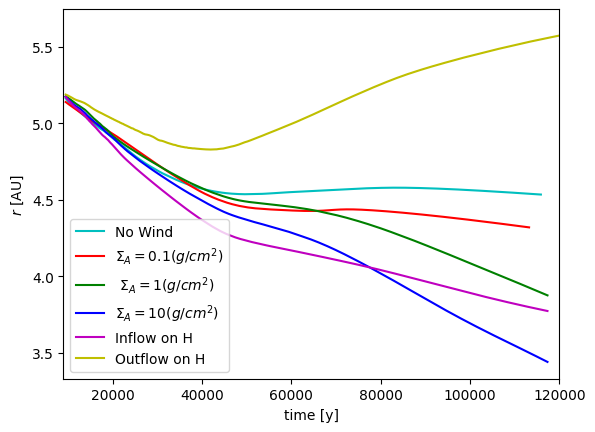}
   \includegraphics[width=9truecm,height=6truecm]{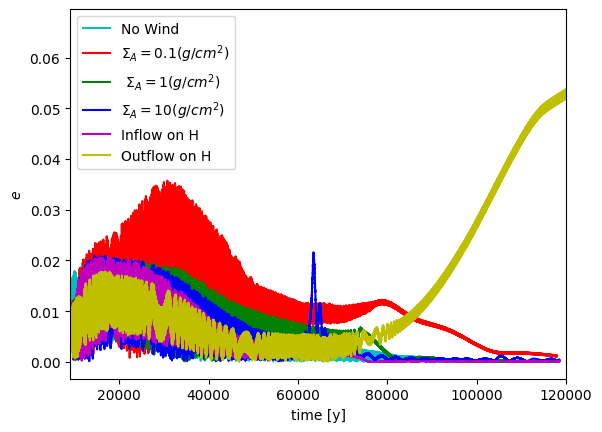}
      \caption{Evolution of the semi-major axes and eccentricities for simulations in layered discs with different column densities of the active layers ($\Sigma_A$). For comparison, data for a classical viscous disc are also plotted (label `No Wind').}
         \label{mig:semiaecc} 
\end{figure}
We report in Fig.\ref{mig:semiaecc} the evolution of semi-major axes and eccentricities for the simulations with transport in surface and midplane layers, and for reference we also plot the case of a classical viscous disc (simulation set "NW").
An initial phase of rapid inward migration is observed for all simulations (up to $t\sim 40000$ y). Eccentricity is slightly excited in this phase and strongly damped later (Fig.\ref{mig:semiaecc}, bottom panel) for all the simulations with the exception of simulation $Hout$ \footnote{In this case, however, eccentricity growth slows towards the end of the simulation, and will be damped on a longer time scale
( similar behaviour was observed in \cite{Legaetal21}
fig.7, bottom panel).}.\par
We have shown in \cite{Legaetal21} that a planet's migration is sustained because the migration of the vortex refills the gap left behind by the migrating planet. We  have called this process `vortex-driven' migration.\par
\begin{figure}
   \includegraphics[width=9truecm,height=7truecm]{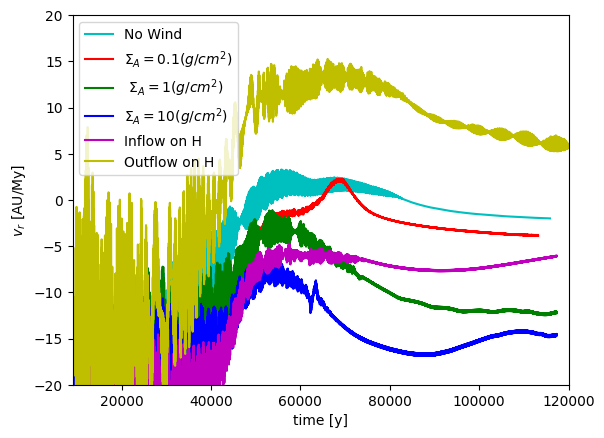}
   \caption{Migration speed vs. time for the six simulation sets.
   A first phase of inward vortex-driven migration is common to all
   the cases. At about $t=50000 yr$ the migration paths bifurcate: for the very thin active layer of simulation $S01$ the final speed of migration is very similar to the $NW$ case; increasing the depth of the layer a regime of faster inward migration is observed.
   When the active layer is at the midplane the planet migrates inwards ($Hin$) and outwards ($Hout$) at opposite speeds in the final phase of the simulation.  }
   \label{mig:speed}
   \end{figure}

In this phase, the synthetic torque generating radial gas flow in the disc's surface layer or midplane layer (up to one disc scale height) appears to have a negligible or moderate effect on the planet's migration. We remark that the inward flow near the midplane (simulation $Hin$) slightly accelerates the planet's inward migration while the outward flow of simulation $Hout$ brakes the vortex-driven inward migration and contributes to an early reversal of the direction of migration.  When the vortex has completely dissipated, the planet's migration appears to
depend on the properties of the layer that transports gas
towards the star (Fig.\ref{mig:semiaecc}, top panel).
For $\Sigma_A=0.1 g/cm^2$, we observe a short phase of outward migration ($70000<t<80000 y$) and finally the planet slowly migrates inwards. The behaviour of this second phase is qualitatively similar to that observed for a planet embedded in a classical, non layered disc with negligible radial flow (the cyan curve in Fig.\ref{mig:semiaecc}, top panel). From Fig.\ref{mig:speed}, however, we see that for $t>80000 \, y$  the planet in simulation $S01$ is migrating inwards somewhat faster than in the classical disc case (simulation set $NW$). In the case of thicker active surface layers,
more rapid inward migration is observed (simulations sets $S1$ and $S10$ in Fig.\ref{mig:semiaecc} and Fig.\ref{mig:speed}).
In the cases $Hin$ and $Hout$ the migration after the dissipation of the vortex is respectively inwards and outwards, like the imposed direction of the gas flow near the midplane  (Fig.\ref{mig:semiaecc} and Fig.\ref{mig:speed}). We explain below the origin of these differences. \par

\par

\begin{figure*}
 \includegraphics[width=8truecm,height=7truecm]{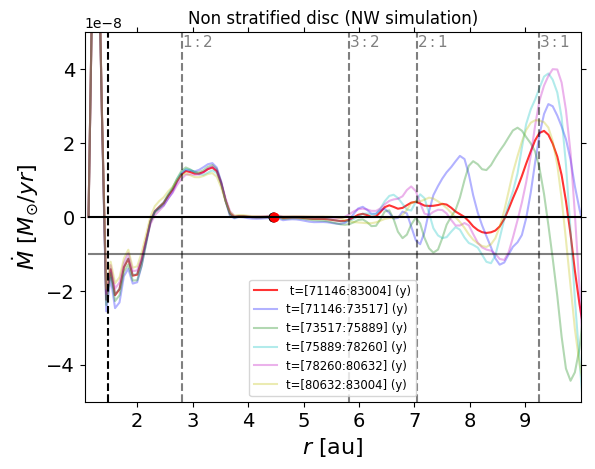}
 \hskip -1.3truecm
   \includegraphics[width=8truecm,height=7truecm]{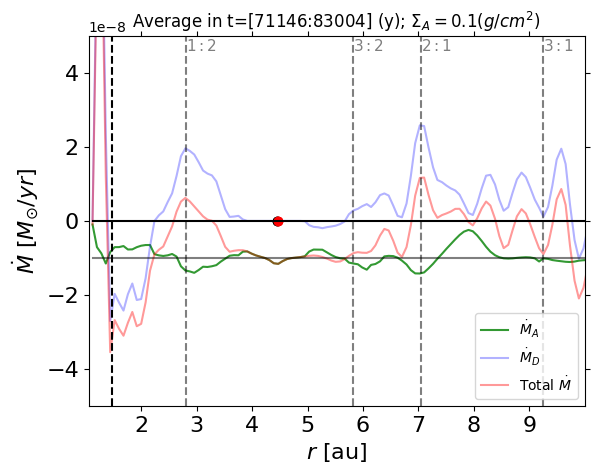}
    \includegraphics[width=8truecm,height=7truecm]{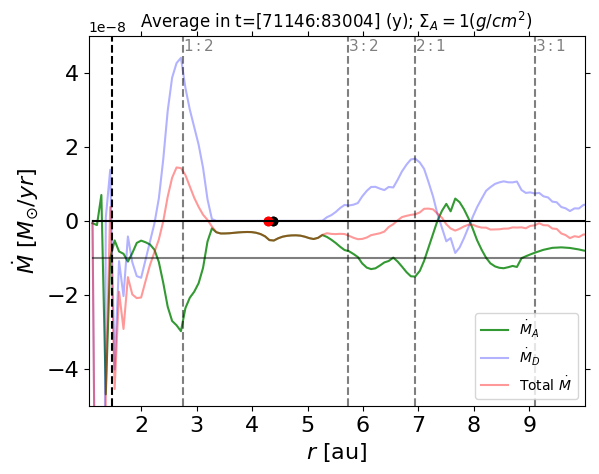}
    %\hskip 1.9truecm
    \includegraphics[width=8truecm,height=7truecm]{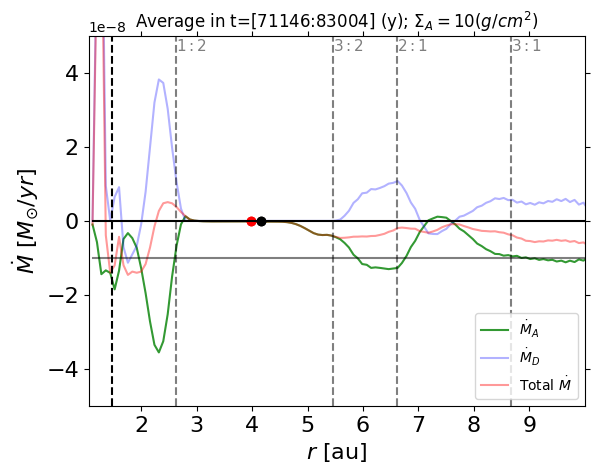} 
    \includegraphics[width=8truecm,height=7truecm]{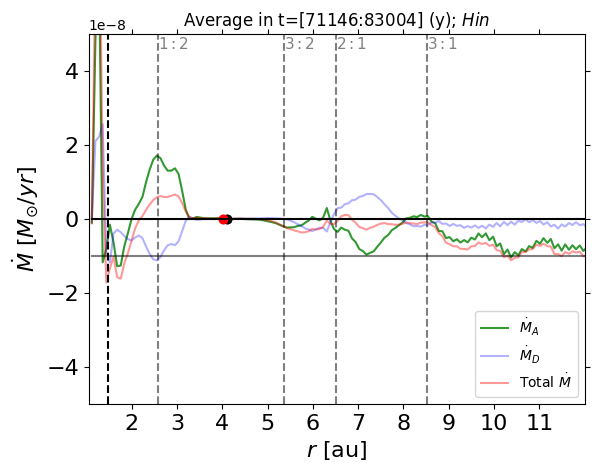}
    \hskip 2.2truecm
   \includegraphics[width=8truecm,height=7truecm]{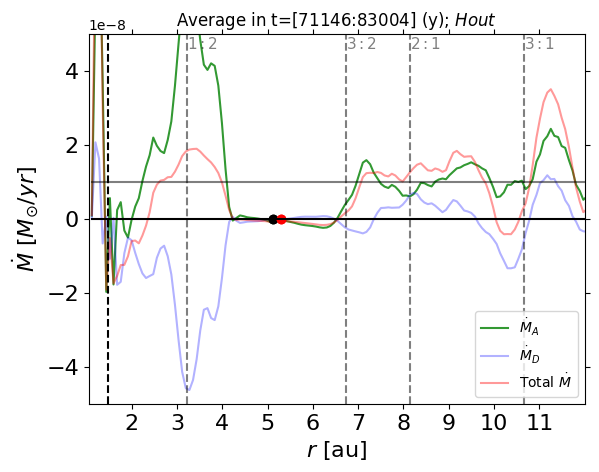}
    \caption{Total mass flow transported through the disc (red curve). A temporal average over 1000 outputs (spaced by $\Delta t = 11.85$ yr) has been performed in order to smooth fluctuations. For the non stratified case (top left panel) we have split the averaging interval into 5 parts. We show the five averages to illustrate the time variations of the flow (see text). In the averaging time interval the planet is  migrating  outwards for simulations $NW$, $S01$ and $Hout$  and is migrating inwards for the cases with thicker active layers: $S1$ and  $S10$ and for  simulation set $Hin$.  The black and red filled circles at $\dot M =0$ indicate the planet-star distance respectively at the beginning and at the end of the averaging interval.
    For the stratified disc models the contribution from the active layers (green curve) as well as the one from the dead region of the disc (blue curve) are also plotted. 
    Notice that for simulations $Hin$ and $Hout$  we apply the synthetic torque in the region between the midplane and $z=H$ (active layers);  while  for $z>H$ no torque is applied (dead layers). The grey horizontal line at  $\dot M = \mp 10^{-8} M_{\sun}/yr$ indicates the expected flow transported by the active layers (the sign plus is for simulation set $Hout$).
     The vertical dotted grey lines provide the location of the main mean motion resonances computed using the planet position at $t=71146 \,y$. The evanescent boundary condition is applied in the domain inside the vertical black dotted line at 1.25 au. }
     \label{mig:radialflow}
    \end{figure*}

\subsection{Radial gas flow through the planet's gap}
\label{sec:RadialFlow}
In order to understand the dependence of the migration speed on
the properties of the active layers, we have computed the net radial gas flow near the region close to the gap (red curves in Fig.\ref{mig:radialflow}). The gas flow has important fluctuations over time; thus, 
to compute the net flow, we performed a temporal average over 1000 orbits in a time interval  centred at about $t=75000$ y.  In this time interval
the planet is very slowly migrating  outwards for simulations $NW$, $S01$ and $Hout$, and is slowly migrating inwards for the other cases (see Fig.\ref{mig:semiaecc}). The migration range is indicated in Fig.\ref{mig:radialflow}: a black and a red filled circles indicate respectively the planet-star distance at the beginning and at the end of the time interval considered for this average. The short migration range justifies the averaging procedure. 
For the stratified disc models the contribution from the active layers (green curve) as well as the one from the non active or dead part   of the disc (blue curve) are also plotted.\par
In the classical non layered disc
(Fig.\ref{mig:radialflow},top left panel) we show the importance of flow fluctuations by reporting also the flow's time average on  five shorter windows of 200 orbits each, with  time intervals  given by: $T_i=[(71146 +(i-1)\times200\times11.85):(71146+i\times200\times11.85)] y$, with $i=1...5$.
In the outer disc ($r>6 $ au) the flow appears to fluctuate significantly over time; the average over 1000 orbits (red curve) appears effective in removing these fluctuations, remaining close to zero except near the $3:1$ mean motion resonance.

\begin{figure}
   \includegraphics[width=7.5truecm,height=7truecm]{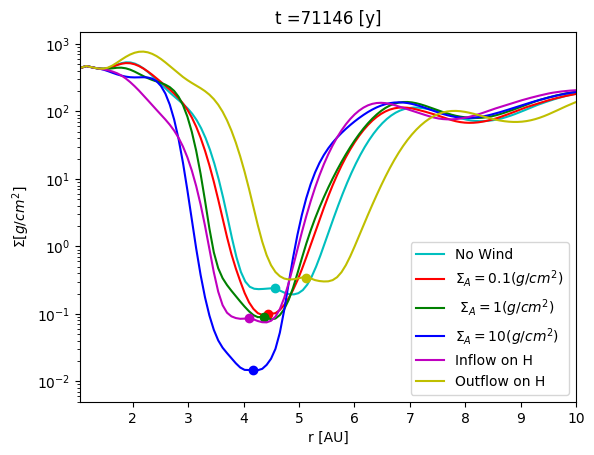}
   \includegraphics[width=7.5truecm,height=7truecm]{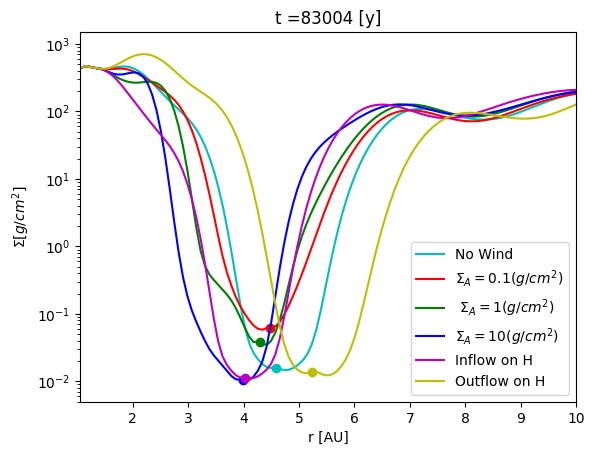}
    \includegraphics[width=7.5truecm,height=7truecm]{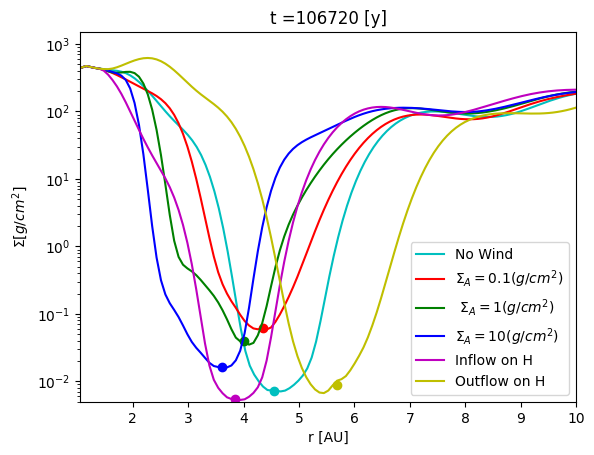}
   \caption{Azimuthally averaged radial surface density profiles  at  specific times (reported on top of the panels). At the time of top and middle panels the migration behaviour has clearly bifurcated with slow outward migration for discs $NW$ and $S01$ and rapid inward migration for the simulation sets with thicker active layers as well as for the case $Hin$. Rapid outward migration is observed for the case $Hout$. 
    Despite the differences in the migration regimes, the gap depth slowly evolves from the time corresponding to the middle panel to the one represented in the bottom panel: gaps are deeper  in all the simulations sets in which the planet blocks the flow from the active layers.}
   \label{mig:sigma2}
   \end{figure}

In the inner disc the flow appears to be stationary and positive at the $1:2$ mean motion resonance. From the observed gas flows, we expect a slowly  pile-up of gas at these resonances (the location of the 1:2, 3:2, 2:1 resonances at $t=71146 y$ are indicated by the dotted  vertical lines in Fig.\ref{mig:radialflow}).  Scaling from our experience with more viscous discs, such resonant perturbations of the flow would require prohibitive long integration times to be averaged out.  What is important to notice in Fig.\ref{mig:radialflow}, top left panel, is that there is no transport of gas across the gap as expected in low viscosity discs (see for example \citet{Robertetal18}).
The shape of the gap at the beginning and at the end of the averaging time interval can be appreciated in Fig.\ref{mig:sigma2}.
\par
In the case of a very thin active layer (Fig.\ref{mig:radialflow}, top right panel) we observe similar fluctuations of the radial flow at the resonant locations in the dead region of the disc (blue curve); instead the radial flow in the active layer (green curve) nicely corresponds to the flow generated by imposing the synthetic torque (Eq.\ref{tqtheo}), which means that the advection of gas in the active layer is barely perturbed by the presence of the planet and its resonances. 
It is worth noticing from Fig.\ref{mig:sigma2} that, for the simulation $S01$, the disc's surface density at the bottom of the gap is of the order of $\Sigma_A = 0.1 g/cm^2$; this means that all the gas in the deepest part of the gap is considered to be 'active'. {  This confirms that} the existence of the gap is not an obstacle to the flow of the active gas. 
\par

In the middle left panel of Fig.\ref{mig:radialflow} (simulation $S1$)
we clearly see that the radial flow transported by the active layers through the gap region is less than half the imposed value $\dot M_A = 10^{-8}$.  This means that the planet acts as a partial barrier for the gas transported by the active layer; therefore gas transported by the active layer piles up at the outer gap's edge. Remember that the planet can migrate only if the disc is able to refill the empty region left behind by the migrating planet; otherwise the planet "detaches" itself from the outer disc and, feeling a progressively weaker negative torque from the outer disc, its migration has to stop eventually. The pile up at the outer edge of the gap of the gas delivered by the active layer is the process that refills the disc and allows the planet to migrate.\par
Increasing further the thickness of the layer (Fig.\ref{mig:radialflow}, middle right panel, simulation set $S10$)  we observe that there is no gas flow through the planet's gap, which means that the planet is able to block the full flow of gas in the active layer and therefore the pile-up effect is even more pronounced. Consequently, inward migration is faster than in the case $S1$. Consistently, we also observe in Fig.~\ref{mig:sigma2} that the density of gas at the bottom of the gap is smaller in the $S10$ simulation than in the $S1$ and $S01$ simulations. In fact, the gas in the gap - which is active - is in steady state and its density is $\Sigma_A f_{gap}$ where $f_{gap}$ is the fraction of the flow of the active layer which is able to penetrate into the gap. The complement of the flow piles up at the outer edge of the gap, at a rate $(1-f_{gap})\dot M_{A}$, driving the planet's migration. {We stress the fact that it is the gas pile up at the outer gap's edge that sustains migration by readjusting the disc so that the planet and the gap can migrate together. Instead, the gas flowing through the gap has no impact on the migration; this flow has an impact only on the gap's depth.}
{\par Precisely, we observe that  the gap depth stabilises and  is very similar in all the simulations with $f_{gap}$ close to zero (middle and bottom panels of  Fig.~\ref{mig:sigma2}) including the reference case without a wind-driven accretion flow ($NW$) and the two simulations with inflow and outflow close to the midplane ($Hin$ and $Hout$). 
As we observe in the bottom panels of Fig.\ref{mig:radialflow} for the two simulations with radial flow imposed in a layer of thickness $H$ at the midplane, the planet appears to be able to block the flow in the gap region. In the case of radial outflow we observe a large effect due to the 1:2 mean motion resonance. Associated with this effect, the surface density profile shows a kink around 3.5 au (Fig. 4). There is also a visible bump at 2 au, a relic of the gap-opening phase (Fig. 3) which, in this case, has not been eroded by inward migration. Both features in the surface density profile are responsible for the positive torque felt by the planet (and hence its outward migration), while the outward radial flow refills the gap left behind by the migrating plane, sustaining the planet migration. This positive torque contribution decreases when the planet migrates further out. Finally,
as previously observed, the migration speed becomes similar 
in magnitude (and opposite in sign)
to the case $Hin$ (Fig.\ref{mig:speed}). In Fig.\ref{mig:radialflow}
bottom left panel, we observe that the contribution of the active layer close to the 1:2 resonance is positive, although the applied synthetic torque provides an inward radial flow as can be appreciated at distances larger than 8 au from the star. We investigate the vertical structure of the flow in the vicinity of the 2:1 resonance in the Appendix (\ref{app:resonance}).
}

  %\subsection{Torque acting on the disc}
  \subsection{Halting the accretion flow and planet migration rates}
  \label{sec:torque}
  In this subsection we consider the conditions under which the torque from the planet can halt the wind-driven accretion flow, and how the planet migration rate is expected to scale with the wind-driven accretion rate through the disc. 
  
   There are two limiting situations to consider. The first corresponds to a very rapid accretion flow in the active layer that can traverse the planet's orbital location unimpeded by the torque from the planet. The active layer has little-to-no effect on the planet's migration rate, which is therefore expected to be close to zero, as in the vortex-driven migration described in \citet{Legaetal21}. This case occurs only if $\Sigma_A$ is sufficiently small because, for a given stellar accretion rate, the radial speed of the flow is inversely proportional to $\Sigma_A$. The other limit corresponds to the planet being able to completely halt the accretion flow, such that no gas crosses the planet's orbit. In this case, as the planet migrates inwards and moves away from the original outer edge of the gap, the accretion flow fills in the gap left behind by the planet and migration proceeds on the time scale over which this occurs. 
  
  %From the analysis of the radial flow we have clearly seen that the planet may act as a barrier to the flow of gas transported in the active layer and that the efficiency of this barrier depends on the thickness of the 
  %active layer. We now analyze why it is so.\par
  
\begin{figure}
   \includegraphics[width=9truecm,height=7truecm]{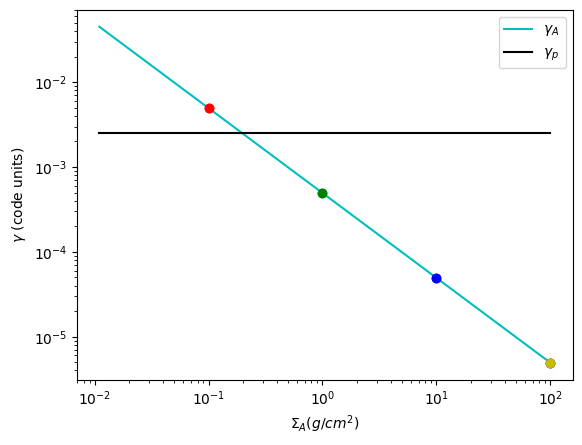}
   \caption{Specific torque $\gamma_A$ and specific planetary torque exerted by a Jupiter mass planet on the outer disc   (Eq.\ref{eq:planettorqueRh}, see text)   $\gamma_p$ as a function of $\Sigma_A$. We notice that  $\gamma_A >\gamma_p$ only for the model with the thinnest layer ($S01$). The dot at $\Sigma_A=100g/cm^2$ refers to the case where the torque is applied to the layer around the midplane ($Hin$ and $Hout$). }
     \label{tqtheo}
   \end{figure}

  First, we compute when the transition should occur between the two limiting cases of unimpeded flow past the planet and no flow past the planet. The torque exerted by the planet on the disc (in our case the active layer) from a distance $r>r_p$ to infinity can be found in \citet{Cridaetal06} and references therein:
  \begin{equation}
       \Gamma_p(\Delta r) = 0.836 q^2 \Sigma_A r_p^4 \Omega_p ^2 \left( r_p \over \Delta_r\right)^3,
      \label{eq:planettorque}
  \end{equation}
  where $\Delta_r= r-r_p$ and $q$ is the ratio between the mass of the planet and the mass of the central star.
  We consider in the following the torque at a distance from the planet equal to the Hill radius, meaning: $\Delta_r = r_p(q/3)^{1/3}$. With this choice, the formula simplifies to:
  \begin{equation}
       \Gamma_p = 3\times 0.836 q \Sigma_A r_p^4\Omega_p^2.
  \label{eq:planettorqueRh}
  \end{equation}
We now consider the specific torque exerted by the planet: 
$\gamma_p= \Gamma_p / (\Sigma_A r^2_p)$
and compare it to the  specific torque $\gamma_A$ 
(Eq.\ref{eq:torquea}) that is imposed in the active layer to generate the
radial flow and associated accretion rate ($\dot M_A = 10^{-8} M_{\sun}/y$ in this study).  
As we see from Eq.~\ref{eq:torquea}, $\gamma_A$  is inversely proportional to  $\Sigma_A$. Fig.~\ref{tqtheo} shows $\gamma_p$ and $\gamma_A$ as a function of $\Sigma_A$, and demonstrates that the synthetic specific torque exerted on the active layer overcomes the specific torque received from the planet only in the case of the thinnest layer (simulation set $S01$) for our assumed accretion rate. This is consistent with the fact that the gas passes through the planet's orbit at a rate that is unaffected by the planet as shown in Fig.~\ref{mig:radialflow}, top right panel.

For the cases $S10$ and $Hin$, $Hout$ (these last two cases have a column density of $\sim 100$g/cm$^2$), however, the torque exerted by the planet overcomes that imposed on the active layer, the accretion flow is completed halted by the planet and gas has to pile up at the outer edge of the gap, thus sustaining migration. The case $S1$ shows significant reduction in the flow past the planet, but not complete blockage of the accretion flow, and so represents an intermediate case between $S01$ and $S10$. 
\par
Equating the expression for $\gamma_p$ with that for $\gamma_A$ from equation~\ref{eq:torquea} gives the column density for the active layer at the transition between impeded and unimpeded accretion flow:
\begin{equation}
    {\tilde \Sigma_A} =  {\dot M_{A} \over 
   12 \times 0.836 \pi q \Omega_p r_p^2 }
   \label{eq:threshold_mp}.
\end{equation}
The previous equation can be rewritten as:
\begin{equation}
{\tilde \Sigma_A}= 0.2{\rm g/cm}^2\,\left(\frac{\dot{M}}{10^{-8}M_\sun/\rm yr}\right)\,\left(\frac{M_p}{M_{\rm Jup}}\right)^{-1}\left(\frac{r_p}{5.2\, \rm au}\right)^{-1/2}\ 
. \label{eq:threshold_mpscale}
\end{equation}
%$$ 
%so that Eq.~\ref{eq:threshold_mp} provides a good estimate of the value of $\Sigma_A$ for which there is a transition between an unimpeded and a completely blocked accretion flow in the active layer.
%However, for an accretion rate onto the star an order of magnitude larger than this, we have ${\tilde \Sigma_A}=2$ g/cm$^2$.  

The equivalence between the two specific torques gives the threshold between the two migration regimes: for $\gamma_p < \gamma_A$ the gas should pass through the planet's orbit without significant pile-up, and the planet is expected to experience only short range migration as in the vortex-driven migration regime. In the opposite limit,  $\gamma_p > \gamma_A$, the flow of gas is inhibited by the presence of the planet and gas piles-up outside of the planet's orbit. The inward migration speed, ${\dot r_{\rm p}}$, is then controlled by the rate at which the accretion flow fills in the gap as the planet migrates inwards. Migration is always the outcome of the imbalance of the torques felt by the planet from the inner and the outer parts of the disc. Because the inner disc preserves its unperturbed density up to the gap's edge, inward migration can be sustained only if the pile-up of gas outside of the gap restores the original unperturbed density $\Sigma$. This leads to the equation: 
\begin{equation}
    2 \pi r_{\rm p} {\dot r_{\rm p}} \Sigma = 2 \pi r_{\rm p} v_{r} \Sigma_{\rm A}, 
    \label{eq:rpdot1}
\end{equation}
which gives the migration rate
\begin{equation}
    {\dot r_{\rm p}} = \frac{\Sigma_{\rm A}}{\Sigma} v_r = \frac{\dot M_{A}}{2 \pi r_{\rm p} \Sigma}.
    \label{eq:rpdot2}
\end{equation}
For a planet orbiting at 5.2 au in a disc with $\Sigma=222$~g/cm$^2$ and $\dot M_{A}=10^{-8}$~M$_{\sun}$/yr, Eq.~\ref{eq:rpdot2} gives $\dot r_{\rm p}= 12.3$~au/y, which is in decent agreement with the migration rate observed in figure~\ref{mig:speed} for run $S10$ with $\Sigma_{\rm A}=10$g/cm$^2$. We expect the migration rate to be slower when the accretion flow is only partially blocked because the rate at which the gap refills is decreased in this situation. This is supported by the fact that run $S1$ produces a slower migration rate than $S10$.

{The  picture presented above neglects second order effects that are likely to be important. One of these is that the migration speed is connected to the asymmetry of the gap associated with the whole disc, and not just to the ability of the planet to block the accretion flow in the active layer. Figure~\ref{mig:sigma2} (middle and bottom panels) shows the bottom of the gap is asymmetric with respect to the planet position in the cases $S1$ and  $S10$, while it is more
symmetric in simulation $Hin$ because in this case $\gamma_p >> \gamma_A$. An effect may also arise from the fact that the accreting layer is near the disc surface in the run $S10$ whereas it is located near the midplane in run $Hin$. These factors likely explain why the final phase of planet inward migration is slower in the case $Hin$ despite the fact that in runs $S10$ and $Hin$ the same net flow is imposed on the disc and 100\% of it is blocked by the torque from the planet.
}
\par
{We also notice that the column density for the active layer at the transition between impeded and unimpeded accretion flow scales with the planet-star distance as  $ r_p^{-1/2}$ (Eq.\ref{eq:threshold_mpscale}).
As a consequence, a planet formed at about 5 au  in a disc with ionised column density  $\Sigma _A$ slightly larger than $\tilde \Sigma _A$ at 5 au (or equivalently  $\gamma _p > \gamma_A$), may experience fast inward migration until it reaches a distance to the star
$\tilde r_p$ such that 
 $\Sigma _A \leq \tilde \Sigma _A(\tilde r_p)$ and the flow of gas is no more inhibited by the presence of the planet.  The planet is then expected to experience  short range migration and possibly stall in the warm-Jupiter region. As an example, in the case of an ionised column density of $\Sigma_A = 0.5 g/cm^2$, and $\dot M_{A}=10^{-8}$~M$_{\sun}$/yr,
 a Jupiter mass planet  is expected to enter in the regime of short range migration at $\tilde r_p \simeq 0.8 $ au.  }\par

Finally in this section, we comment on how interactions between gap opening planets with classical viscous discs differ compared to interactions with laminar discs that have layered accretion. In the  model of a layered disc we have considered, the torque per unit mass acting on the accreting material is set by the mass accretion rate and the column density of the active layer, and if these do not change then the torque has a constant and well defined value. A planet in this case can block the accretion flow by exerting the required torque on the active layer, as discussed above. In the case of a viscous disc, however, the formation of a gap generates radial gradients that increase the rate of diffusion of gas into the gap. This is illustrated by the equation for the radial velocity in a viscous disc
\begin{equation}
    v_R = \left[\frac{\partial (R^2 \Omega)}{\partial R} \right]^{-1} \frac{1}{R \Sigma}\frac{\partial}{\partial R} \left(R^3 \nu \Sigma \frac{d \Omega}{dR}\right),
\end{equation}
which shows the radial velocity increases in magnitude in response to gradients developing in the flow. Thus, in the viscous case a gap is carved until an equilibrium is eventually  reached between the torque exerted by the planet on the disc and the viscous torque (setting the gap's radial profile); instead, in the case we consider in this paper no equilibrium between $\gamma_p$ and $\gamma_A$  occurs and the gas either piles up far enough from the planet's orbit or passes through the planet's orbit basically unimpeded.
{ We recall that we consider  the wind torque to be steady with a prescribed efficiency, neglecting  possible inhomogeneities arising from a radial transport of magnetic flux in the gap's vicinity. Therefore, in our model the radial flow due to the wind torque is not associated with an increase of $\alpha$ viscosity in the gap. A radial flux of angular momentum, inducing a radial flow like in an alpha viscous model, requires the modelling of the magnetic field and is expected to depend on its geometry \citep{2013A&A...550A..61L}}.\par
One consequence of { the fact that the gas either piles-up or passes quasi unimpeded through the planet's orbit}  is that we might expect the accretion of gas onto giant planets to occur at different rates, depending on whether or not they are embedded in viscous or layered discs, even when the global mass accretion rate through the disc is the same. This issue will be explored in a forthcoming paper (Nelson et al., in preparation).

%This explains why simulations of Jovian mass planets embedded in traditional viscous disc models have been shown to accrete at a rate of $10^{-5}$~M$_{\rm Jup}$/yr $\equiv 10^{-8}$~M$_{\sun}$/yr, in spite of opening a deep gap \citep{Bryden1999,Kley1999,Lubow1999}. 

\section{Migration of a Saturn mass planet}
   \label{sec:Saturn}
   From the analysis of the migration of a Jupiter mass planet we have concluded that the  equivalence between the two specific torques gives the   
   threshold layer depth $\tilde \Sigma_A $ that separates the regime of short range migration from the one of fast inward migration.
   To confirm our findings,
   in this section we study the migration of a Saturn mass planet in discs with active layers that have different column densities. 
%   In order to choose the depth of the active layers corresponding to the two migration regimes we first compute 
%    $\tilde \Sigma_A $  
%from the equivalence between Eq.\ref{eq:planettorque} and Eq.\ref{eq:torquea} :
%\begin{equation}
 %  \tilde \Sigma_A = {\dot M_{A} \over 
  % 12 \, 0.836 \pi q \Omega_p r_p^2 } 
   %\label{eq:threshold_mp}
%\end{equation}

We test the validity of Eq.~\ref{eq:threshold_mp} by simulating the migration of a Saturn-mass planet in discs with column layers of depth $\Sigma_A=0.3$~g/cm$^2$ and $\Sigma_A=1$~g/cm$^2$. These two values bracket the value of ${\tilde \Sigma_A} = 0.65$~g/cm$^2$ obtained from Eq.~\ref{eq:threshold_mp} for a Saturn mass planet and ${\dot M_A} = 10^{-8}$~M$_{\odot}$/yr. Additionally we run a reference simulation
for the disc without a wind-driven accretion layer. 
We use the same prescription as in Eq.\ref{eq:massgrowth} for the growth of the planet to its final mass value.  We report in Table \ref{table:tab2} the names and the main simulation parameters.\par

%(Fig.\ref{mig:surfdensS}) we observe in the
% left panels that a vortex is formed at the outer gap edge. When comparing to Fig.\ref{mig:surface} we see that by reducing the mass of the planet the strength of the vortex appears to weaken.
% The vortex spreads and practically disappears
% (rightmost panels of Fig.\ref{mig:surfdensS}),
% with a temporal sequence very similar to the one
% observed in the previous case (Fig.\ref{mig:surface}).
% Qualitatively, the dynamics of the vortex and 
% of planet migration is unchanged with respect to the case of the more massive planet; we inspect quantitatively the migration paths and the radial flow behaviour in the following.}
 \par
\begin{figure}
   \includegraphics[width=8truecm,height=6truecm]{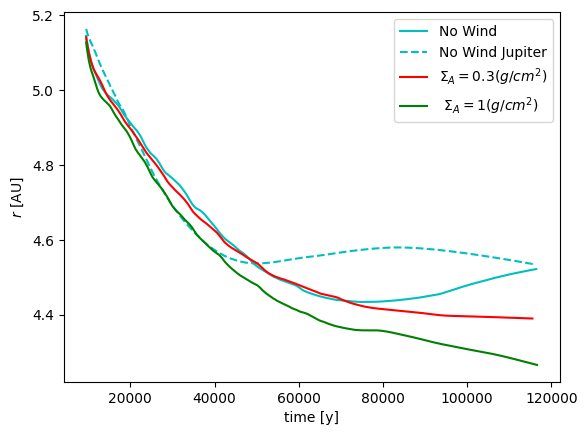}
      \caption{Evolution with time of semi-major axes  for simulations in layered discs with different depths of the active layer ($\Sigma_A$) for the case of a Saturn mass planet. For comparison we also plotted (cyan dashed line) the data for simulation $NW$.}
         \label{mig:semiaeccS} 
\end{figure}

 
 It is useful to start from the reference case $NW_S$ for which we observe a phase of rapid inward migration (Fig.\ref{mig:semiaeccS}) followed by a phase of slower outward migration after vortex dissipation. The mechanism is the same as described for the Jupiter planet case, however the range of inward migration is slightly larger (for comparison we also plotted in Fig.\ref{mig:semiaeccS}  the data corresponding to simulation $NW$). By reducing the mass of the planet the strength of the vortex appears to weaken. A weaker vortex  opens a less deep secondary gap and the balance between the torques from the inner and the outer disc occurs later in time, when the planet is slightly closer to the star with respect to the more massive planet case. The planet migrates outwards in the final phase of the run with speed slightly reducing towards the end of the simulation.
 \par
 The migration of the planet in the two layered discs confirms the expected 
 behaviour: in both cases we observe "vortex-driven" migration and the migration paths bifurcate after vortex dissipation: in the case of the thin layer (simulation set $S03_S$) the planet's migration slows down and practically stops towards the end of the simulation,
 while for the thicker layer, after a very short phase of outward migration, the planet migrates inwards again.
 No eccentricity excitation is observed in the three cases (not shown here).
 \begin{table}
      % Give a unique label
%\vbox to100mm{\vfil
% For LaTeX tables use
%\begin{center}
%\begin{minipage}{100mm}
\begin{tabular}{|llll|}
\hline\noalign{\smallskip}
Name & $\Sigma_A (g/cm^2)$ & ${\dot M_A (M_{\odot}/y)}$ & ${\dot M_{D} (M_{\odot}/y)}$ \\
\noalign{\smallskip}\hline\noalign{\smallskip}
\hline\noalign{\smallskip}
$NW_S$  & 0  &  $0$ & $-8\,10^{-11}$ \\
$S03_S$  & 0.3  &  $-10^{-8}$ & $-8\,10^{-11}$ \\
$S1_S$  & 1  &  $-10^{-8}$ & $-8\,10^{-11}$  \\
\noalign{\smallskip}\hline
\end{tabular}
\caption{Simulation names and main parameters for the case of a Saturn mass planet. The accretion  flow transported by the active layers is labelled $\dot M_A$,  while $\dot M_{D}$ is the star accretion rate due to viscosity. The minus sign indicates inflow.}
\label{table:tab2} 
\end{table}

We confirm therefore that the migration behaviour after vortex 
dissipation depends on the magnitude of   the specific torque exerted by the planet at the outer edge of the gap with respect to the torque exerted by the wind. \par

  \section{Conclusions and discussion} 
  \label{sec:Conclusion}
   The migration of giant planets in classical low viscosity discs has been recently studied by \citet{Legaetal21}. Interestingly, in discs with vanishing viscosity migration is driven by the migration of the vortex that forms at the outer edge of the gap. Under some conditions this vortex-driven migration can be very slow and then reverses, until it eventually stops. This result is very promising to explain the limited migration that most giant planets seem to have experienced, as is the case for the "cold-Jupiter" population. However, classical low viscosity discs do not account for the gas accretion rates observed onto young stars. \par
    In this paper we have studied giant planet migration in low viscosity discs in which 
  we have modelled the radial transport of gas in laminar accretion layers. This radial transport mimics the effect of angular momentum loss due to magnetised discs winds, or due to horizontal fields near the midplane being generated by the Hall effect, and account for the typically observed gas accretion rates of ${\dot M} = 10^{-8}$~M$_{\odot}$/yr.
  We have modelled the effect of the disc wind by applying a synthetic torque in a surface layer of the disc characterised by a prescribed column density $\Sigma_A$.
 We have also considered a case with accretion focussed near the disc midplane to mimic transport properties of disc outer regions characterised by weak Ohmic diffusion or due to the Hall effect.
 
Migration appears to be characterised by two phases. In the first phase, planet migration is driven by the vortex and is directed inwards.
This phase ends when the vortex disappears after having opened a secondary gap, as typically observed in vortex-driven migration \citep{Legaetal21}.  In this phase, the speed of migration is independent of the thickness of the active layer. In the second phase, migration is observed to depend on the thickness of the active layer. \par
By comparing the torque exerted by the planet to the prescribed torque applied in the active layers
we have shown that gas advection   modifies  planet migration when the torque exerted by the planet  overcomes the torque applied in the active layers. 
In this case the gas transported by the active layer piles up at the outer gap's edge. 
Considering that giant planet migration is possible  only if the disc is able to refill the empty region left behind by the migrating planet,  the gas pile up at the outer edge of the gap is the process that refills the gap.\par
On the contrary, when the torque applied on the active layers overcomes the torque exerted by the planet the gas transported by the active layers is not blocked by the planet and there is no pile up of gas to sustain fast inward migration.  A slow inward migration is observed, similar to that seen in simulations of low viscosity discs without laminar accretion flows \citep{Legaetal21}.\par
We remark that our result is different from the one of \citet{McNallyetal2020}, who found that planet migration was not affected by the presence of an accreting layer near the disc surface, but this is not surprising since the cited study concerned fully embedded low mass planets.  
For Jupiter-mass planets and a stellar accretion rate of $10^{-8} M_{\sun}/y$, the separation between the two regimes, for a planet at 5.2 au occurs for a value of the  vertically integrated column density of the active layer of about 0.2~g/cm$^2$, while in the case of
a Saturn mass planet this happens at 
about 0.65~g/cm$^2$.
{ We notice, however, that the column
density giving the separation between the two regimes scales
with the planet-star distance as $r_p^{-1/2}$ so that a fast migrating
giant planet should eventually enter in the regime of slow inwards
migration once it is close enough to the star. The large number of giant
planets in the warm-Jupiter region suggests that the column density of
the active layer is of the order of  0.2 - 0.5 g/cm$^2$ , which would stall a
Jupiter-mass planet at $\sim 4 - 1$ au respectively.}

It is noteworthy that although x-rays and cosmic rays are important sources of ionisation in protoplanetary discs, and penetrate into columns of depth $\sim 10$~g/cm$^2$ and 100~g/cm$^2$, respectively, the strong magnetic coupling in the surface layers of protoplanetary discs that allows a magnetised wind to be launched is thought to depend on the ionsation of sulphur and carbon atoms by UV photons from the star \citep{PerezBeckerChiang2011}. The column density of this ionised layer has been calculated to be in the range 0.01 to 0.1 g/cm$^2$. For  this range of values we expect Jupiter mass planets stalling in the Warm Jupiter region.
\par
Our model is certainly simplistic with respect to a full non-ideal MHD model. However it has its merits: on the one hand a full MHD model with an embedded planet evolving on long timescales is not yet accessible with available computational resources; on the other hand our model  captures the essential interaction of the planet with a layer torqued by some external process, and hence provides a framework with which to understand future more complex simulations when they become possible. From the results we have obtained, we expect that giant planets embedded in low viscosity stratified discs may have a rich and still not fully explored variety of dynamical paths, depending on the properties of the active layers. Among these paths there is vortex-driven migration \citep{Legaetal21}, which is extremely slow after a short-ranged phase, and therefore has the potential to explain the orbits of cold Jupiters (at a few au from the central star) even if these planets formed no farther than the snowline location (5-10 au).

\begin{acknowledgements}
 We acknowledge support by DFG-ANR supported GEPARD project
 (ANR-18-CE92-0044 DFG: KL 650/31-1). We also acknowledge HPC resources from GENCI DARI n.A0100407233 and from "Mesocentre SIGAMM" hosted by Observatoire de la C\^ote d'Azur. LE wish to thank Alain Miniussi for maintenance and re-factorisation of the code FARGOCA. RPN acknowledges support from STFC through the consolidated grants ST/M001202/1 and ST/P000592/1.
\end{acknowledgements}

% WARNING
%-------------------------------------------------------------------
% Please note that we have included the references to the file aa.dem in
% order to compile it, but we ask you to:
%
% - use BibTeX with the regular commands:
%   \bibliographystyle{aa} % style aa.bst
%   \bibliography{Yourfile} % your references Yourfile.bib
%
% - join the .bib files when you upload your source files
%-------------------------------------------------------------------

% WARNING
%-------------------------------------------------------------------
% Please note that we have included the references to the file aa.dem in
% order to compile it, but we ask you to:
%
% - use BibTeX with the regular commands:
%   \bibliographystyle{aa} % style aa.bst
%   \bibliography{Yourfile} % your references Yourfile.bib
%
% - join the .bib files when you upload your source files
%-------------------------------------------------------------------

\begin{appendix}

\section{Three dimensional investigation of the 1:2 mean motion resonance}
\label{app:resonance}
\begin{figure}
 \includegraphics[width=8truecm,height=12truecm]{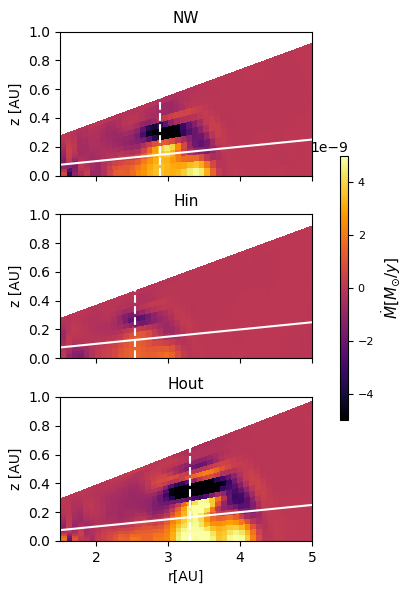}
    \caption{Azimuthally averaged radial flow in the plane $r-z$ at time $t=83004$ yr for simulations sets $NW$, $Hin$ and $Hout$.  In simulations $Hin$ and $Hout$ a synthetic torque is applied below the disc scale height indicated  in the figure by the white continous line. The dashed vertical line indicates the position of the 1:2 mean motion resonance considering planet position at time $t=83004$ yr.}
    \label{app:RZflow}
    \end{figure}

The radial flow analysis that we provided in the paper is a one dimensional analysis that provides useful information. However, in order to have a better insight on the flow behaviour, and specifically to analyse the most persistent resonant effect that we have observed in all the simulations presented in the paper, we provide in this appendix a two dimensional ($r-z$) view of the radial flow in the vicinity of the 1:2 mean motion resonance. Fig.\ref{app:RZflow}
shows the azimuthally averaged radial flow for simulations $NW$, $Hin$ and $Hout$. 
An important amount of outflow is observed mainly up to $z=h_0r$, while the flow is strongly inwards in a thin layer located above $z>h_0r$. When comparing simulation $Hin$ and $Hout$ to the reference simulation $NW$ we see that the torque imposed up to a disc scale height respectively decreases and enhance the natural outflow at the 1:2 resonance, according to the direction of the imposed flow. The imposed inflow, in simulation $Hin$, appears not enough to overcome the natural outflow associated to the 1:2 resonance and the resulting total flow is still positive as observed in Fig.\ref{mig:radialflow}, bottom left panel.
The imposed outflow of in simulation $Hout$, on the contrary, contributes in enhancing the gas motion at the resonant location (Fig.\ref{mig:radialflow}, bottom right panel). 

\end{appendix}
\end{document}